# μSR study of the Cu-spin dynamics in the electron-doped high-$T_c$ cuprate of $Pr_{0.86}LaCe_{0.14}Cu_{1-y}(Zn,Ni)_yO_4$

Risdiana[a*], T. Adachi[a], Y. Koike[a] and I. Watanabe[b]

[a]Department of Applied Physics, Tohoku University, 6-6-05, Aoba, Aramaki, Aoba-ku, Sendai 980-8579, Japan
[b]Advanced Meson Science Laboratory, RIKEN, Wako 351-0198, Japan

**Abstract**

Effects of the Zn- and Ni-substitution on the Cu-spin dynamics in the electron-doped $Pr_{0.86}LaCe_{0.14}Cu_{1-y}(Zn,Ni)_yO_{4+\alpha-\delta}$ with y = 0, 0.01, 0.02, 0.05 and different values of the reduced oxygen content $\delta$ have been studied using zero-field muon-spin-relaxation (μSR) measurements at temperatures down to 2 K. For the as-grown sample ($\delta = 0$, y = 0) and the sample with a very small $\delta$ value ($\delta < 0.01$, y = 0), a muon-spin precession due to long-range antiferromagnetic order has been observed. On the other hand, no precession has been observed for moderately oxygen-reduced samples ($0.01 \leqq \delta \leqq 0.09$). It has been found that for all the samples of $0.01 \leqq \delta \leqq 0.09$ the asymmetry A(t) (μSR time spectrum) in the long-time region increases with decreasing temperature at low temperatures, suggesting possible slowing-down of the Cu-spin fluctuations. On the other hand, no significant difference between Zn- and Ni-substitution effects on the slowing down of the Cu-spin fluctuations has been observed.

## 1. Introduction

The study of the impurity effect, namely, the effect of the partial substitution of Zn or Ni for Cu on the Cu-spin dynamics in the high-$T_c$ cuprates has attracted much attention in relation to the mechanism of superconductivity. A lot of experimental results on the impurity effect have been reported for the hole-doped high-$T_c$ cuprates [1-3], in contrast to those for the electron-doped ones. The difficulty in preparing superconducting samples of good quality in the electron-doped cuprates is one of the reasons. In the electron-doped cuprate with the so-called T' structure, superconducting samples are obtained after the heat treatment of as-grown samples in a reducing atmosphere [4]. The superconducting properties such as the superconducting transition temperature, $T_c$, are affected by the reduced oxygen content, $\delta$, as well as by the impurity concentration [5].

Here, we have investigated the Zn- and Ni-substitution effects on the Cu-spin dynamics in the electron-doped $Pr_{0.86}LaCe_{0.14}Cu_{1-y}(Zn,Ni)_yO_{4+\alpha-\delta}$ from muon-spin relaxation (μSR) measurements, changing y up to 0.05 and $\delta$ up to 0.09 [6].

## 2. Experimental

Polycrystalline samples of $Pr_{0.86}LaCe_{0.14}Cu_{1-y}(Zn,Ni)_yO_{4+\alpha-\delta}$ with y = 0, 0.01, 0.02 and 0.05 were prepared by the ordinary solid-state reaction method from $La_2O_3$, $Pr_6O_{11}$, $CeO_2$, CuO and ZnO or NiO powders of high purity [6,7]. As-grown samples were post-annealed in flowing Ar gas of high purity (6N) at 950 °C for 10 h in order to remove the excess oxygen. The $\delta$ value was estimated from the weight change before and after post-annealing. To check the quality of the obtained samples and to determine $T_c$, both electrical resistivity and magnetic susceptibility were measured. Zero-field (ZF) μSR measurements were performed at temperatures down to 2 K at the RIKEN-RAL Muon Facility at the Rutherford-Appleton Laboratory in UK.

---

[*] Corresponding author. Tel.: +81-22-795-7977; Fax: +81-22-795-7975; e-mail: risdiana@teion.apph.tohoku.ac.jp

## 3. Results and discussion

The obtained μSR time spectra of impurity-free $Pr_{0.86}LaCe_{0.14}CuO_{4+\alpha-\delta}$ are roughly grouped into 4 classes with different $\delta$ values: as-grown ($\delta = 0$), very small $\delta$ ($\delta < 0.01$), small $\delta$ ($0.01 \leq \delta < 0.04$), and large $\delta$ ($0.04 \leq \delta \leq 0.09$). Samples with the small and large $\delta$ show superconductivity with $T_c$ ranging from 15 K to 17 K (average $T_c \sim 16$ K) and from 18 K to 22 K (average $T_c \sim 20$ K), respectively, while the as-grown sample and the sample with very small $\delta$ are not superconducting above 4.2 K.

Figure 1 shows typical ZF-μSR time spectra of impurity-free $Pr_{0.86}LaCe_{0.14}CuO_{4+\alpha-\delta}$. For the as-grown sample, a muon-spin precession is observed even at high temperature of 100 K due to long-range antiferromagnetic order. For the samples with very small $\delta$, small $\delta$ and large $\delta$, a Gaussian-like behavior is observed at high temperatures above ~ 100 K due to randomly oriented nuclear spins, and an exponential-like depolarization of muon spins is observed at low temperatures below ~ 50 K. For the sample with very small $\delta$, a muon-spin precession is observed at low temperatures below ~ 5 K. For the samples with small and large δ, on the other hand, no muon-spin precession is observed, indicating the absence of any long-range magnetic order above 2 K. The temperature-dependent change of the spectra above 20 K is regarded as being due to the static random magnetism of small $Pr^{3+}$ moments [8]. Concerning the samples with small and large δ (0.01 ≦ δ ≦ 0.09), it is found that the asymmetry A(t), namely, the μSR time spectrum in the long-time region around 10 μsec, increases with decreasing temperature at low temperatures, suggesting possible slowing-down of the Cu-spin fluctuations.

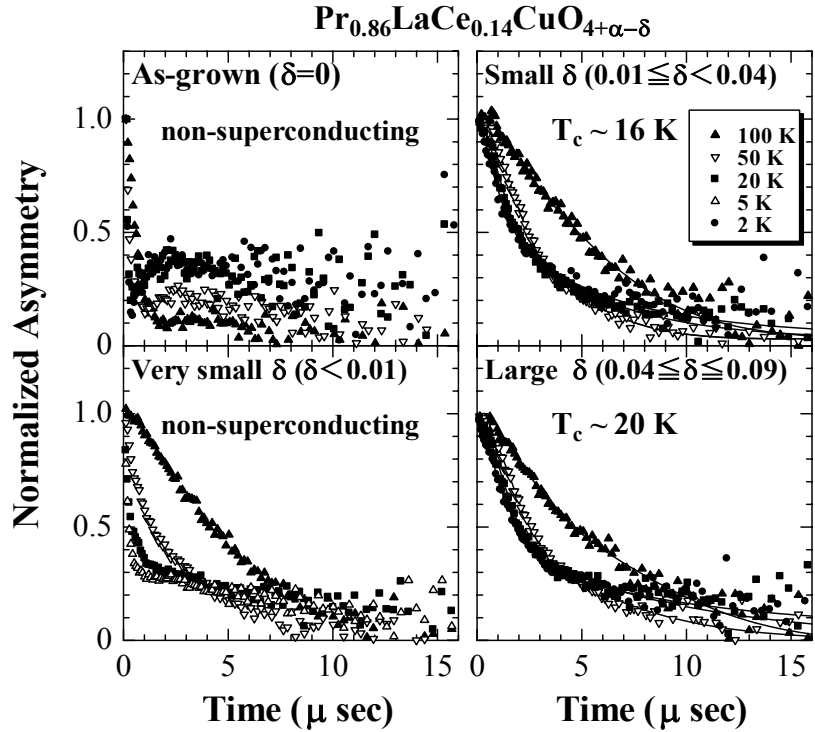

**FIGURE 1.** Typical ZF-μSR time spectra of $Pr_{0.86}LaCe_{0.14}CuO_{4+\alpha-\delta}$ with different $\delta$ values at various temperatures. Solid lines for the small and large $\delta$ are the best-fit results using a two-component function: $A(t) = A_G\exp[-\sigma^2 t^2]+A_s\exp[-(\lambda t)^\beta]$.

The ZF-μSR time spectra are analyzed with the following two-component function: $A(t) = A_G \exp[-\sigma^2 t^2] + A_s \exp[-(\lambda t)^\beta]$. The first term is a static Gaussian component in the region where the relaxation due to nuclear spins and small $Pr^{3+}$ moments [8] is dominant. The second term is a dynamical stretched-exponential component in the region where the Cu-spin fluctuations exhibit slowing down. The increase of A(t) in the long-time region at low temperatures is reflected by the increase in $A_s$. The time spectra are well fitted with this function as shown in Fig. 1.

Figure 2 shows the temperature dependence of the fitted parameter $A_s$ for samples with various y and δ values in the range 0.01 ≦ δ ≦ 0.09. For the impurity-free sample of y = 0 with large δ, for example, it is found that $A_s$ decreases with decreasing temperature down to 30 K, which is due to the growing effect of the $Pr^{3+}$ moments. Below ~ 30 K, $A_s$ increases with decreasing temperature, indicating the slowing down of the Cu-spin fluctuations. For the impurity-free sample with small $\delta$, on the other hand, $A_s$ increases with decreasing temperature below 45 K. The difference of the temperature where $A_s$ shows the minimum may be due to the residual effect of a small amount of antiferromagnetically ordered Cu-spins in the impurity-free sample with small $\delta$. In any case, it attracts interest that the slowing down of the Cu-spin fluctuations is observed even in the impurity-free sample. This may be due to possible enhancement of the Cu-spin correlation assisted by the $Pr^{3+}$ moments.

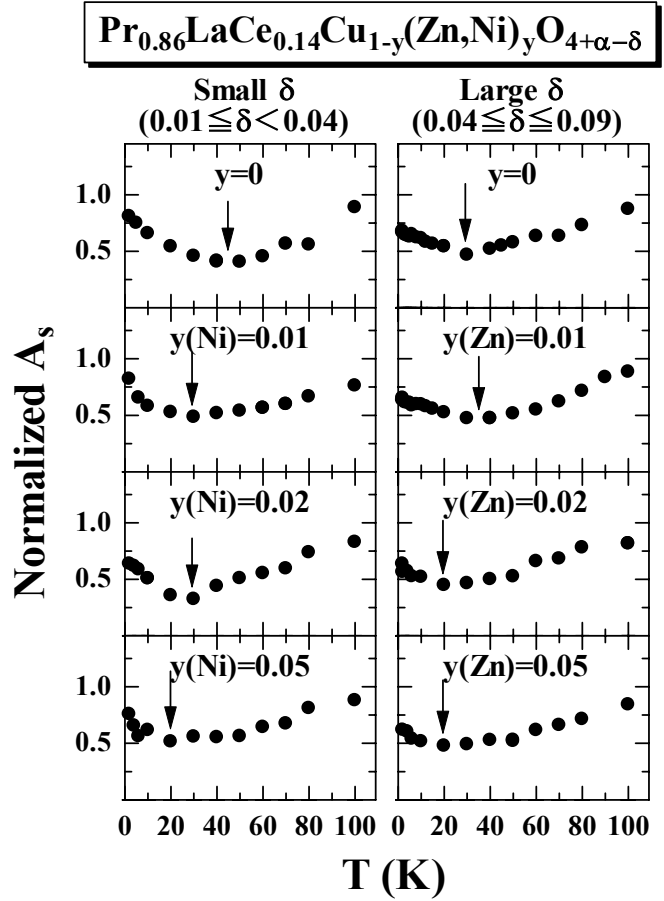

**FIGURE 2.** Temperature dependence of $A_s$ for $Pr_{0.86}LaCe_{0.14}Cu_{1-y}(Zn,Ni)_yO_{4+\alpha-\delta}$ with y = 0, 0.01, 0.02, 0.05 and small and large δ values at temperatures down to 2 K. Arrows indicate the temperature where $A_s$ exhibits the minimum.

The increase in $A_s$ at low temperatures is still observed for both Zn- and Ni-substituted samples up to y = 0.05. However, no significant difference in the temperature dependence of $A_s$ between Zn- and Ni-substituted samples is observed. It appears that the effect of the $Pr^{3+}$ moments is stronger than that of a small amount of Zn and Ni impurities. These behaviors are very different from those observed in the hole-doped high-$T_c$ cuprates [2,3].

## 4. Summary

We have investigated the Zn- and Ni- substitution effects on the Cu-spin dynamics from ZF-μSR measurements in the electron-doped cuprate $Pr_{0.86}LaCe_{0.14}Cu_{1-y}(Zn,Ni)_yO_{4+\alpha-\delta}$ with y = 0, 0.01, 0.02, 0.05 and various $\delta$ values at temperatures down to 2 K. It has been found that the Cu-spin fluctuations exhibit slowing down at low temperatures in both impurity-free and impurity-substituted samples, regardless of the y value for moderately oxygen-reduced samples ($0.01 \leqq \delta \leqq 0.09$). A possible origin of the slowing down observed even in the impurity-free sample is enhancement of the Cu-spin correlation assisted by the $Pr^{3+}$ moments. No significant difference in the temperature dependence of $A_s$ between Zn- and Ni-substituted samples may be due to the stronger effect of the $Pr^{3+}$ moments than that of a small amount of Zn and Ni impurities.

## Acknowledgments

This work was supported by Joint Programs of the Japan Society for the Promotion of Science, TORAY Science and Technology Grant and also Grant-in-Aid for Scientific Research from the Ministry of Education, Culture, Sports, Science and Technology, Japan.